\documentclass[12pt]{article}
\usepackage{amssymb}
\usepackage{amsmath}
\begin{document}
\noindent
{\Large {\bf Stochastic integrability and the KPZ
equation}}\bigskip\\
{\large by H}ERBERT {\large S}POHN {\large (Munich, Germany)}\bigskip\bigskip\\
\par
\begingroup
\leftskip=5.2cm 
\noindent{\small\textit{Herbert Spohn received his Ph.D. in physics at the
Ludwig-Maximi-lians-Universit\"{a}t, M\"{u}nchen. He is now professor
for Mathematical Physics at the Zentrum Mathematik, Technical
University Munich, with joint appointment by the Physics Department.
His main research focus is nonequilibrium statistical mechanics.
He has published ``Large Scale Dynamics of Interacting
Particles" at Springer-Verlag and ``Dynamics of Charged Particles
and Their Radiation Field" at Cambridge University Press.}}
\par
\endgroup
\noindent
{\small\textit{Spohn was awarded  the 2011 Dannie
Heineman Prize for Mathematical Physics, the 2011 Leonard Eisenbud
Prize for Mathematics and Physics, the 2011 Caterina Tomassoni
Prize, and a Ph.D. honoris causa of Universit\'{e} Paris-Dauphine.
He served in the IAMP executive committee 1997--1999 and as president
2000--2002.}}\bigskip\\
   As a common experience from basic courses in Classical Mechanics,
for some mechanical systems the equations of motion can be solved up to
quadratures, while others persist to deny such access. This
experience can be formalized and leads to the notion of an
integrable system. For a Hamiltonian system with $n$ degrees of
freedom, one requires to have at least $n$ functions on phase space,
$H_j$, $j=1,\ldots,n$, which are in involution meaning that the Poisson brackets $\{H_i,H_j\}=0$ for
$i,j=1,\ldots,n$,  see \cite{1} for details. $H_1$, say, is the system's Hamiltonian. 
Then the manifold $\{H_j=c_j$, $j=1,\ldots,n\}$
has the structure of an $n$-torus and the motion is characterized by
at most $n$ frequencies. Hence, up to deformation, the motion looks
like the well-known Lissajous figures.

The text book example is the motion of a particle subject to a
central potential. More spectacular is the observation that
integrability persists for particular  systems with a large number of
degrees of freedom, which first surfaced indirectly through the
discovery of solitary wave solutions by N.J. Zabusky and M.D.
Kruskal \cite{2} for the Korteweg-de-Vries equation in one dimension
and for a chain of nonlinear coupled oscillators by M. Toda
\cite{3}. A very rich field ensued. In the following my focus will be on the aspect of
many interacting components.

Naturally one may wonder how integrability survives under
quantization. An Hamiltonian operator, $H$, allows for many commuting
operators. Thus a simple minded extension from the classical case
will not do and there seems to be no generally agreed upon definition of
quantum integrability. On the other side there are clear signatures
to identify a quantum integrable system (once it is found), to name
only a few: Bethe ansatz, Yang-Baxter equation, and factorized $S$-matrix.

From the perspective of statistical mechanics it is also a natural issue to
understand whether and how integrability extends to stochastic
systems. To have one distinction very clear, many systems of 2D
equilibrium statistical mechanics are integrable, the
correspondence being related to the fact that the transfer matrix has a
structure akin to a quantum integrable system. In contrast, here I
discuss stochastic time evolutions modeled as a Markov process,
either diffusion or jump. As a linear operator, the generator, $L$, of the
Markov process has possibly some structural similarity to $-H$,
hence it seems reasonable to expect a corresponding version of integrability. On the other
side, $\mathrm{e}^{Lt}$ is already the normalized transition
probability; there are no probability amplitudes, the partition function equals 1, and the 
largest real part of the eigenvalues is 0.

With R. Dobrushin the Russian probability school pioneered the many
component aspect. Integrability is usually first associated  with R. Glauber's
exact solution of the one-dimensional stochastic Ising model
\cite{4}. This solution is based on what is now called duality, a concept
introduced and generalized to other systems by F. Spitzer in the
very influential article \cite{5}. The dual description is here in terms
of evolution equations for the time-dependent correlations functions, which decouple
for integrable systems. An example is the
symmetric simple exclusion process on the one-dimensional lattice
$\mathbb{Z}$. In this model there is at most one particle per site
and, under this restriction, particles perform independently nearest
neighbor symmetric random walks. The generator $L$ equals $-H$ with
$H$ the Hamiltonian of the ferromagnetic Heisenberg chain. (In this
case, duality holds in arbitrary dimension and also for longer
ranged symmetric jumps.)

On the level of duality, none of the signatures known for quantum
integrability make their appearance. This situation changes
drastically as we turn to the asymmetric version of the simple
exclusion process, ASEP (now 1D and n.n. do matter). A particle at site
$j$ jumps to site $j+1$ with rate $p$ and to site $j-1$ with rate
$q$, $q+p=1$, provided the destination site happens to be empty. The
symmetric case corresponds to $q=p=\frac{1}{2}$. The generator can
be written in the notation of quantum spin chains. If $\sigma^z_j=1$
means site $j$ is occupied by a particle, then
\begin{equation}\label{1}
L=\frac{1}{4}\sum_{j\in\mathbb{Z}}
\big(\overset{\rightharpoonup}{\sigma}_j\cdot\overset{\rightharpoonup}{\sigma}_{j+1}
-1 + 2i (p-q)(\sigma^x_j \sigma^y_{j+1} -\sigma^y_j
\sigma^x_{j+1})\big)\,.
\end{equation}
Note that $L$ is not symmetric. All eigenvalues are in the open left
hand plane except for 0. On a ring with a fixed number of particles, the 
unique invariant measure is the uniform distribution.  The
other eigenvectors are determined through the Bethe ansatz \cite{6}.
Much more powerful is the Bethe ansatz inspired expression for the
transition probability $\mathrm{e}^{Lt}$ discovered by C. Tracy an
H. Widom \cite{7}. Their expression is still extremely complicated
and to simplify further one has to specify some initial conditions.
A widely studied choice is the initial step, for which the half
lattice $\{j\leq 0\}$ is empty and $\{j\geq 1\}$ occupied. For
$q>p$ Tracy and Widom write a Fredholm determinant for the probability distribution of
$x_j(t)$, the position of the $j$-th particle
at time $t$. Much earlier K. Johansson \cite{8} found  a distinct Fredholm determinant
for a related quantity in the totally
asymmetric limit $q=1$ (TASEP). Both results serve as the stepping stone for
an intricate asymptotic analysis eventually arriving at objects familiar from
random matrix theory.

Very recently one accomplished to cross the border from discrete jump
processes to a particular stochastic PDE, which reads
\begin{equation}\label{2}
\frac{\partial}{\partial t}h=\frac{1}{2}\big(\frac{\partial}{\partial
x} h\big)^2 + \frac{1}{2}\frac{\partial^2}{\partial x ^2} h + W
\,,\quad x\in \mathbb{R}\,, \,t\geq 0\,,
\end{equation}
and is the 1D version of the equation first proposed by Kardar, Parisi, and Zhang
\cite{9} as a model for growing interfaces. Here $h(x,t)$ is viewed as a height
function and $W(x,t)$ is white noise in space-time. Integrability is
seen most explicitly for the sharp wedge  initial condition,\begin{equation}\label{3}
h(x,0)=-\frac{1}{\delta} |x|\quad \textrm{with}\; \delta \to 0\,,
\end{equation}
which, at $\delta = 1$, should be understood as the analogue of the once integrated initial step. (\ref{2}) together with (\ref{3}) looks very singular, and it is.
For smooth initial data the solution is constructed by L. Bertini
and G. Giacomin \cite{10} and for the sharp wedge in \cite{11}.

The KPZ equation turns linear through the Cole-Hopf
transformation
\begin{equation}\label{4}
    Z=\mathrm{e}^h\,.
\end{equation}
Then
\begin{equation}\label{5}
\frac{\partial}{\partial t}Z=\frac{1}{2}\frac{\partial^2}{\partial x ^2}  Z + W Z\,,\quad Z(x,0) = \delta(x)\,,
\end{equation}
from which one concludes that the exponential moments of $h$ are
linked to the attractive $\delta$-Bose gas in one dimension, which is a quantum 
integrable system solvable through the Bethe ansatz. For
example, for (\ref{2}) together with (\ref{3}) it holds
\begin{equation}\label{6}
\mathbb{E}(Z(0,t)^n)= \langle 0|\mathrm{e}^{-t H_n} |0
\rangle
\end{equation}
with $H_n$ the $n$ particle attractive Lieb-Liniger hamiltonian,
\begin{equation}\label{7}
H_n=-\frac{1}{2}\sum^n_{j=1} \frac{\partial^2}{\partial
x_j^2}-\frac{1}{2}\sum^n_{i\neq j=1}\delta(x_i-x_j)\,,
\end{equation}
and $|0\rangle$ the state where all $n$ quantum particles are at 0.
Unfortunately, the moments in (\ref{6}) increase as $\exp(n^3)$,
which makes a rigorous control difficult. But replica schemes have
been employed and yield fascinating results \cite{12,12a,13,14,15}.

Currently the integrability of the KPZ equation can be deduced only
indirectly by taking a continuum limit of the asymmetric simple
exclusion process, where the lattice spacing is $\varepsilon$, the
time scale $\varepsilon^{-2}$, and the asymmetry
$q-p=\sqrt{\varepsilon}$ with $\varepsilon\ll 1$. To give an
impression, I record the generating function for the height at
the origin at time $t$, 
\begin{equation}\label{8}
\mathbb{E}\big(\exp \big[-\mathrm{e}^{-s}
\mathrm{e}^{h(t)+(t/24)}\big]\big)= \det (1-P_0 K_{s,t} P_0)\,.
\end{equation}
Here the determinant is in $L^2(\mathbb{R})$, $P_0$ projects onto
$[0,\infty)$, and $K_{s,t}$ is an operator with integral
kernel
\begin{equation}\label{9}
K_{s,t}(x,y)= \int_\mathbb{R}\big(1+\mathrm{e}^{-(t/2)^{1/3}\lambda
+s}\big)^{-1}\mathrm{Ai}(x +\lambda) \mathrm{Ai}(y+\lambda) \mathrm{d}\lambda
\end{equation}
with Ai the Airy function. $P_0 K_{s,t} P_0$ is of trace-class. For large $t$, $h(t)\cong -t/24+(t/2)^{1/3}
\xi$, where the random amplitude $\xi$ is Tracy-Widom distributed, just
as is the largest eigenvalue of a GUE random matrix in the large $N$
limit. (\ref{8}) together with (\ref{9}) was obtained independently in
\cite{11,16,16a}. In this context the introductory review \cite{17} is
highly recommended with some complementary information provided in
\cite{18}.

The integrability of the KPZ equation triggered further advances.
One interesting direction is to consider discretized versions of
the stochastic heat equation (\ref{5}). Somewhat unexpectedly, the
completely asymmetric discretization turns out to be more tractable and one
starts from the equations of motion
\begin{equation}\label{10}
\frac{d}{dt} Z_j(t)=Z_{j-1}(t) - Z_{j}(t) +\big(\tfrac{d}{dt} b_j(t)\big) Z_j(t)\,,\quad
Z_j(0)=\delta_{j0}\,,\;j\in \mathbb{Z}\,,
\end{equation}
where $\{b_j(t),j\in\mathbb{Z}\}$ is a collection of independent,
standard Brownian motions. N. O'Connell \cite{19} established a close
connection between $\log Z_n(t)$ and the last particle in the open
quantum Toda chain of $n$ sites. Very recently A. Borodin and I.
Corwin \cite{20} explain how Macdonald functions enter the picture. 
They are the eigenfunctions of the commuting set of Macdonald operators.
In the future, for sure, the interface between stochastic and
quantum integrability will be further elucidated.

While we emphasized the notion of integrability, let me add as a fairly extended
footnote that the predictions based on the exact solutions have been
confirmed recently  in spectacular experiments \cite{21}, see also the expository article
\cite{22}. Of course, physical systems are much more complex than
simple models as the TASEP. But on a large space-time scale
microscopic details hardly matter, except for generic properties,
like the condition of a sufficiently local update rule. In fact,
such universal behavior can be proved for the integrable models discussed, but
it should hold at much greater generality, including physical
systems. In the experiment \cite{21} one studies droplet growth
in a thin film of turbulent liquid crystal. The film thickness is 12~$\mu$m, while the droplet 
grows laterally to a size of  several mm. The
droplet consists of the stable DSM2 phase and is embedded in the
metastable DSM1 phase. Hence the interface is a line and it advances through
nucleation events where the stable phase is created out of  the metastable one. On
average, the
solution to the KPZ equation with sharp wedge initial data has a parabolic profile which self-similarly widens linearly in
$t$ and thus models one section of the droplet. By the physical conditions, the
droplet growth is isotropic guaranteeing that the non-universal
coefficients do not depend on the direction of growth, which is the
basis for high precision sampling of entire probability density functions. In fact,
the GUE Tracy-Widom distribution for the height fluctuations is
confirmed with accuracy. It is also observed that for flat
initial conditions, $h(x,0)=0$, the height fluctuations switch from
GUE to GOE statistics, implying that some features of the initial
conditions are still visible in the large scale universal limit.\\\\


\textbf{References}\vspace{-1.5cm}

\begin{thebibliography}{10}
{\footnotesize 

\bibitem{1} V.I. Arnold, Mathematical Methods of Classical Mechanics. Springer-Verlag, New York, 1978.

\bibitem{2} N.J. Zabusky, M.D. Kruskal, Interaction of "solitons" in a collisionless plasma and the recurrence of initial states, Phys. Rev. Lett. \textbf{15}, 240--243 (1965).

\bibitem{3} M. Toda, Vibration of a chain with nonlinear interaction, J. Phys. Soc. Japan  \textbf{22}, 431--436 (1967).

\bibitem{4} R.J. Glauber, Time-dependent statistics of the Ising model, J. Math. Phys.  \textbf{4}, 294--308 (1963).

\bibitem{5} F. Spitzer, Interaction of Markov processes, Advances in Math. \textbf{5},  246--290 (1970).

\bibitem{6}  Leh-Hun Gwa, H. Spohn, Bethe solution for the
dynamical scaling exponent of the noisy Burgers equation,
Phys. Rev. A {\bf 46}, 844--854 (1992).

\bibitem{7} C.A. Tracy and H. Widom, Integral formulas for the asymmetric simple exclusion process,
Comm. Math. Phys. \textbf{279},  815--844   (2009).

\bibitem{8} K. Johansson, Shape fluctuations and random matrices, Comm. Math.
Phys. \textbf{209}, 437--476 (2000).

\bibitem{9} M.~Kardar, G.~Parisi, Y.C. Zhang, Dynamic scaling of
growing interfaces, Phys. Rev. Lett. \textbf{56}, 889--892 (1986).

\bibitem{10} L. Bertini, G. Giacomin, Stochastic Burgers and KPZ equations from particle systems,
Comm. Math. Phys. \textbf{183}, 571--607 (1997).

\bibitem{11} G. Amir, I. Corwin, J. Quastel, Probability distribution of the
free energy of the continuum directed random polymer in $1+1$
dimensions, Comm. Pure Appl. Math. \textbf{64}, 466--537 (2011).

\bibitem{12} P. Calabrese, P. Le Doussal, A. Rosso,
Free-energy distribution of the directed polymer at high
temperature, EPL \textbf{90}, 20002 (2010).

\bibitem{12a} V. Dotsenko,
Bethe ansatz derivation of the Tracy-Widom distribution for
one-dimensional directed polymers,   EPL \textbf{90}, 20003 (2010).

\bibitem{13} P. Calabrese, P. Le Doussal, Exact solution for the Kardar-Parisi-Zhang equation with flat
initial conditions, Phys. Rev. Lett. \textbf{106}, 250603 (2011).


\bibitem{14} T. Imamura, T. Sasamoto, Exact solution for the stationary KPZ equation,
\texttt{arXiv:1111:4634}.

\bibitem{15} S. Prolhac, H. Spohn, The one-dimensional KPZ
equation and the Airy process, J. Stat. Mech. (2011) P03020.

\bibitem{16} T. Sasamoto, H. Spohn, One-dimensional Kadar-Parisi-Zhang
equation: an exact solution and its universality, Phys. Rev. Lett.
\textbf{104}, 230602 (2010).

\bibitem{16a} T. Sasamoto and H. Spohn, Exact height distributions for the
KPZ equation with narrow wedge initial condition, Nuclear Phys. B
\textbf{834}, 523--542 (2010).

\bibitem{17}  I. Corwin, The Kardar-Parisi-Zhang equation and universality class,
\texttt{arXiv:1106.1596}.

\bibitem{18} T. Sasamoto, H. Spohn, The 1+1-dimensional Kardar-Parisi-Zhang
equation and its universality class, Proceedings StatPhys 24, J.
Stat. Mech. (2011) P01031.

\bibitem{19} N. O'Connell, Directed polymers and the quantum Toda lattice, Ann. Prob., to appear (2012).

\bibitem{20} A. Borodin, I. Corwin, Macdonald processes, \texttt{arXiv:1111.4408}.

\bibitem{21}K.A. Takeuchi, M. Sano, Evidence for geometry-dependent universal fluctuations of the Kardar-Parisi-Zhang interfaces in liquid-crystal turbulence, \texttt{arXiv:1203.2530}.

\bibitem{22} K. Takeuchi, M. Sano, T. Sasamoto, H. Spohn,
Growing interfaces uncover universal fluctuations behind scale
invariance, Scientific Reports \textbf{1}, 34 (2011).}


\end{thebibliography}
\end{document}